\documentstyle[12pt,aaspp4]{article}
\begin{document} 
\title{Potential Sources of Gravitational Wave Emission and
Laser Beam Interferometers }
\author{J.A.de Freitas Pacheco}
\affil{Observatoire de la C\^ote d'Azur, B.P. 229, F-06304
 Nice Cedex 4, France \\ 
e-mail: pacheco@obs-nice.fr }
\date{Received date; accepted date}

\begin{abstract}

The properties of potential gravitational wave sources like
neutron stars, black holes and binary systems are reviewed, as well as
the different contributions (stochastic and continuous) to the gravitational 
wave background. The detectability of these sources by the present generation
of laser beam interferometers, which will be fully operational around 2002, is 
also considered.
\end{abstract} 

\bigskip
\bigskip
\noindent
{\bf Paper presented at the 6$^{th}$ International Workshop on Relativistic
 Aspects of Nuclear Physics  - Tabatinga - Brazil - October 2000}

\newpage

\section{Introduction}

Gravitational waves (GW) are predicted by the General Relativity Theory (GRT) and the
slow inspiral observed in the binary pulsar  PSR 1913+16 (Hulse \& Taylor 1975)
is an indirect proof of their  existence.  GW are fundamentally different from 
electromagnetic waves. While the latter  propagate in  the framework of space and 
time, the former are waves of the spacetime itself, created by asymmetric mass 
motions. A direct detection of GW has not been achieved up  to date.

A resonant-mass antenna is, in principle, the simplest detector of GW (Weber 1960).
A suspended cylindric detector, during the passage of a gravitational wave, has its 
normal modes excited and monitored by transducers. Present day resonant  bars  (or
''spheres") are fairly narrowband detectors, with bandwidths of only a few Hz  around
frequencies in the kHz range and sensitivities around  {\bf h} $\approx$ 10$^{-19}$.
Future detectors should have bandwidths  of the order of 100 Hz or better and
sensitivities of about 10$^{-21}$.

In the early seventies emerged the idea that laser interferometers might have a better
chance of detecting GW (Weiss 1971; Moss, Miller \& Forward 1971). These
detectors are essentially constituted by a two-arm Michelson interferometer, which
measures the phase difference between a splitted laser beam having propagated
along two perpendicular directions. This is the quantity that would be changed by
a passing and properly oriented gravitational wave. Laser beam interferometers
are wideband detectors, being sensitive to GW in the frequency range 10-10$^4$ Hz.
 Plans for kilometer-size
interferometers have been developed in the past decades. The US project {\bf LIGO}
is under development at two widely separated  sites (Hanford and Livingston), both
localities hosting a 4 km interferometer. The 3 km French-Italian antenna  {\bf VIRGO} is
being built in Cascina, near Pisa, and a sophisticated seismic isolation system 
will allow this
detector to measure frequencies down to 5 Hz. {\bf GEO 600} is a 0.6 km arm 
interferometer
in construction by  a British-German collaboration, in a site near Hannover. Presently, the 
only operational laser interferometer is {\bf TAMA}  (0.3 km arms), located in Mitaka (Japan).
Besides these ground based antennas, there is also a project supported by NASA and ESA
to launch a large spatial interferometer (5 million km arms), constituted by three
platforms. The {\bf LISA} antenna  will search for low-frequency (mHz) GW
sources, which cannot be observed from the ground because of tectonic activity.   

The best signal-to-noise (S/N) ratio that can be achieved from these detectors
implies the use of ''matched-filter'' techniques, that require a priori
knowledge of the waveform. Thus, in this context, the study of the most probable
GW sources is of fundamental importance. In this work, the properties of some 
GW sources that have been discussed in the literature in the past years will 
be reviewed, as well as the expected 
detectability by the major interferometers under development in the world.

\section{Neutron Stars}

\subsection{Rotating neutron stars - pulsars}

Rotating neutron stars may  have a time-varying quadrupole moment and hence 
radiate GW, by either having a triaxial shape or a misalignment between the
symmetry and the spin axes, which produces a wobble in the stellar motion
(Ferrari \& Ruffini 1969; Zimmerman \& Szednits 1979). In the former case
the GW frequency is equal to twice the rotation frequency, whereas in the
latter  two modes are possible: one in which the GW  have the same frequency
as the rotation, and another in which the GW have twice  the rotation frequency.
The first mode dominates by far at small wobble angles while the importance
of the second increases for large values of the misalignment.

In the case of a rotating triaxial neutron star, the gravitational strain amplitude of
both polarization modes are:
\begin{equation}
h_{+}(t) = 2A(1 + \cos^2i)\cos(2\Omega t)
\end{equation}
and
\begin{equation}
h_{X}(t) = 4A\cos i \sin(2\Omega t)
\end{equation}
where {\bf i} is the angle between the spin axis and the wave propagation vector,
assumed to coincide with the line of sight, $\Omega$ is the angular rotation 
velocity of the neutron star,
\begin{equation}
A = {{G}\over{rc^4}}\varepsilon I_{zz}\Omega^2
\end{equation}
G is the gravitation constant, c is the velocity of light, r is the distance to the
source  and the ellipticity $\varepsilon$ is defined as
\begin{equation}
\varepsilon = {{I_{xx} - I_{yy}}\over{I_{zz}}}
\end{equation}
with  $I_{ij}$ being the principal inertia moments of the star.

\begin{equation}
h(t) = h_+(t)F_+(\theta ,\phi ,\psi) + h_{\times}(t)F_{\times}(\theta ,\phi 
,\psi)
\end{equation}

where $F_+$ and $F_X$ are the beam factors of the interferometer, which
are functions of the zenith distance $\theta$, the azimuth $\phi$ as well as
of the wave polarization plane orientation $\psi$ (see, for instance, Jaranowski
et al. 1998, for details).

Detection of both gravitational polarization modes of a radio pulsar leads immediately
to the value of the spin projection angle i  and  to an estimate of the
ellipticity, if the distance is known by measuring the dispersion of 
radio signals through the interstellar plasma. Upper limits of $\varepsilon$
have been obtained by assuming that the observed spin-down of pulsars is
essentially due to the emission of GW. In this case, for  ''normal'' pulsars one
obtains $\varepsilon \leq 10^{-3}$ whereas  ''recycled'' millisecond pulsars
seem to have equatorial deformations less than
$10^{-8}$. Monte Carlo simulations of the galactic pulsar population
by Regimbau \& de Freitas Pacheco (2000a) indicate that with the planned VIRGO
sensitivity (better than LIGO at lower frequencies) and integration times
of the order of $10^7$ s, a few detections should be possible if $\varepsilon$
= $10^{-6}$. 

What are the physical mechanisms able to deform the
star ? Different scenarios leading to a distorted star
have been discussed in the literature, as anisotropic stresses from strong
magnetic fields, and tilting of the symmetry axis during the initial cooling
phase when the crust solidifies. Bildsten (1998) pointed out that a neutron 
star in a state of accretion
may develop non-axisymmetric temperature variations in the surface, which
produce horizontal density patterns able to create a mass quadrupole
moment of the order of  $10^{38}$  g.cm$^{-2}$, if the elastic response of 
the crust is neglected. More detailed calculations by Ushomirsky et al. (2000)
indicate that the inclusion of the crustal elasticity decreases by a factor 20-50
the expected mass quadrupole moment. Nevertheless, the deformation induced by 
the accretion process is able to balance the angular momentum gained by mass
transfer to that lost by GW emission, imposing limits to the maximum
rotation frequency attained by the  ''recycled'' neutron star. Strong  magnetic
fields are also able to distort the star (Gunn \& Ostriker 1970; Bonazzola
\& Gourgoulhon 1996). Recent calculations by Konno et al. (2000) including
general relativity corrections permit to estimate the ellipticity by the relation
\begin{equation}
\varepsilon_B \approx 4\times 10^{-8}g B_{14}^2
\end{equation}
where g is a parameter depending on the structure of the neutron star, in particular
on the adopted equation of state, and $B_{14}$ is the magnetic field strength in units
of  10$^{14}$ Gauss. According to those authors, a typical value of the structure
factor is g $\approx$ 14. In this case, 
if an ellipticity of $10^{-6}$ is required, then from the above
equation a field of 1.3$\times 10^{14}$ G is derived. The number of pulsars
in the Galaxy with magnetic fields higher than $10^{14}$ may be considerable,
about 23\% of the total population, if the magnetic field decay timescale is
much longer than the pulsar lifetime (Regimbau \& de Freitas Pacheco 2000b). However
these objects can be discarded as potential GW sources since they are rapidly 
decelerated by magnetic dipole  emission and, consequently, most of them have 
presently periods higher than 20 s. 

\subsection{Bar-mode instabilities}

Rotating and self-gravitating incompressible fluids are subjected to non-axisymmetric 
instabilities when the ratio $\beta$ =  T/$\mid W\mid$ of the rotational energy T
to the gravitational energy W is sufficiently large (Chandrasekhar 1969). These
instabilities correspond to global nonradial toroidal modes with eigenfunctions $\propto$
$e^{\pm im\phi}$, where
m = 2 is the  so-called  ''bar'' mode, the fastest growing mode when rotation is
very rapid. Incompressible Newtonian stars in the presence of some dissipative
mechanism (viscosity  or gravitational radiation reaction) become {\it secularly}
unstable against bar formation when $\beta \geq$ 0.14.  In this case, the instability growth 
is essentially determined by the shortest dissipative timescale. On the other  hand, when
$\beta \geq$ 0.27, the star becomes {\it dynamically} unstable to bar formation,  and
the growth of the instability is determined by the hydrodynamical timescale of the system.
These instability limits have rigorously been derived for homogeneous and uniformly
rotating Newtonian stars, but further relativistic numerical studies using polytropic 
equations of state and assuming ad-hoc rotation velocity profiles concluded that
the onset of the instabilities occurs approximately at the same limits (see, for
instance, Shibata et al. 2000). 

Numerical simulations indicate that bar formation in dynamically unstable stars is
accompanied by mass and angular momentum losses (Houser et al. 1994; Lai \&
Shapiro 1995), with the ejected matter  forming spiral arms in the equatorial plane.
The subsequent evolution is rather uncertain. Some simulations  suggest that the bar
shape is short lived, while other simulations  predict a  lifetime of many bar-rotation
periods (see, for instance, New et al. 1999). The angular momentum carried out by
the ejected mass and by gravitational waves reduces $\beta$ to values  below the 
dynamical instability limit, but still above the secular stability limit. The resulting
axisymmetric system may evolve into a nonaxisymmetric configuration on a
secular dissipation timescale (Lai \& Shapiro 1995), as we shall see later.     

Estimates of the characteristic gravitational strain amplitude for the dynamical 
instability phase ($\beta \geq$ 0.27) performed by different authors (Houser et 
al. 1994; Houser \& Centrela 1996; Brown 2000; Shibata et al. 2000) seem to 
be in rough agreement, namely, they predict a characteristic strain amplitude 
$h_c \approx$ 3$\times 10^{-22}$ at a distance
of 20 Mpc. However the characteristic frequency of the waves differs 
considerably from author to author,
ranging from 0.49 kHz (Brown 2000) up to 4.0 kHz (Houser et al. 1994).
Adopting the low  frequency estimate a signal-to-noise ratio S/N $\approx$ 2.0
is obtained if
the planned VIRGO sensitivity  curve is used ( a similar result is obtained for
LIGO), whereas  a S/N ratio one order of magnitude smaller is expected
if the characteristic frequency is higher than 2-3 kHz.

Most of the numerical studies  suggest that after the dynamical instability
phase (if the system was initially set beyond the limit $\beta_{crit} \approx$ 0.27),
it recovers almost an axisymmetric shape, but with $\beta$ still above the
secular instability threshold. In this case, the system may evolve away from
the axisymmetric configuration, in a timescale determined by the gravitational
radiation reaction, which is of the order of few seconds for $\beta$ in the range
0.20-0.25, as simulations suggest. This evolutionary path is possible if gravitation radiation
reaction overcomes viscosity. Then during the evolution the fluid circulation
is conserved (but not angular momentum) and the system evolves  toward
a Dedekind ellipsoid, whose configuration is a fixed triaxial figure with an
internal fluid circulation of constant vorticity. In the opposite situation, when
viscosity drives the instability, angular momentum is conserved (but not
the fluid circulation) and the system evolves toward a Jacobi ellipsoid. 

The transition to a Dedekind configuration manifests in the form of strong
hydrodynamic waves in the outer layers  and mantle, propagating
in the opposite direction of the star's rotation. According to Lai \& Shapiro 
(1995), the frequency of the GW is an increasing function of the angular
velocity of the star. The frequency of the waves is maximum at the beginning
of the transition ($\nu_{max} \sim$ 800 Hz) and then it decreases
monotonically. Since GW carry away the star angular momentum, the
final configuration is a nonrotating triaxial ellipsoid, which no more emits
GW. Thus, the wave amplitude first increases, reaches a maximum (when
$\nu \sim$ 500 Hz) and then decreases to zero again. 
Lai \& Shapiro (1995) estimated that the total
gravitational wave energy radiated during the transition could be as large as
$4\times 10^{-3}$Mc$^2$. This is much larger than the expected energy
($\sim 10^{-8}Mc^2$) radiated in the axisymmetric collapse and bounce preceding
the neutron star formation (Zwerger \& Muller 1997). Using the wave amplitude
estimated by Lai \& Shapiro (1995), the expected S/N ratios for  sources at
a distance of 20 Mpc are  4.0 and 3.0 for VIRGO  and LIGO respectively.

\subsection{R-modes in rotating neutron stars}

The r-mode (r for rotation) is a member of the class of gravitational radiation 
driven instabilities (including the secular bar-mode instability) excited
by the so-called CFS (Chandrasekhar-Friedman-Schutz) mechanism (see a recent review by
Andersson \& Kokkotas 2000). The criterion for 
triggering the instability is quite simple: if the pattern speed of the mode is forward-going 
as seen from a distant observer, but backward-going with respect to the rotation of the
star, then when the disturbance radiates away the star angular momentum, the system can
find a state of lower energy and angular momentum. These large scale toroidal fluid
oscillations are similar to the well known geophysical Rossby waves and, in both
cases, the restoring force is the Coriolis force. 

In the first tens of seconds after the formation of the neutron star, the temperature is very
high (T$\approx$ 10$^{10}$ K) and the instabilities discussed in the preceding section
(bar-modes) are thought to be at work. If the temperature is still higher, the bulk viscosity is 
expected to suppress the CFS instability, whereas the shear viscosity plays a stabilizing
role for temperatures T$\leq$ 10$^7$ K. Thus, there is a well defined window in which
the  ''newly-born'' neutron star is unstable. In the plane $\Omega \times$T, the instability
limit region can roughly be estimated by imposing that the total energy rate of the mode
is zero, or equivalently 
\begin{equation}
{{1}\over{\tau_{gr}}} + {{1}\over{\tau_b}} + {{1}\over{\tau_s}} = 0
\end{equation}
where $\tau_{gr}$, $\tau_b$ and $\tau_s$ are respectively the gravitation radiation, bulk
and shear viscosity damping timescales. For the m=2 mode, these timescales are approximately
given by
\begin{equation}
{{1}\over{\tau_{gr}}} = - 0.303({{\Omega^2}\over{\pi G\rho}})^3  \,\  s^{-1}
\end{equation}
\begin{equation}
{{1}\over{\tau_b}} = 5\times 10^{-12}T_9^6({{\Omega^2}\over{\pi G\rho}}) \,\  s^{-1}
\end{equation} 
and
\begin{equation}
{{1}\over{\tau_s}} = 3.3\times 10^{-9}T_9^{-2}  \,\  s^{-1}
\end{equation}
where T$_9$ is the temperature in units of 10$^9$ K, $\Omega$ is the angular velocity
of the star and $\rho$ its mean density. As the star cools by neutrino emission 
and decelerates by GW emission, it stabilizes around periods of 15-25 ms. According 
to this scenario, no ''newly-born'' pulsar faster than this limit should be 
observed. However, these estimates depend not only on damping effects due to 
different physical mechanisms, most of which are still badly understood, but 
also on the role played by the crust and magnetic field.

For purposes of detection, the most important properties of the GW signal from r-modes
are: a) the frequencies,  which at the lowest order in the angular velocity of the star are given 
 by
\begin{equation}
\omega_r = {{(m-1)(m+2)}\over{(m+1)}}\Omega
\end{equation}
Thus, when m = 2, the GW frequency is ${\nu_{gr}} = {{4}\over{3}}\nu_{rot}$;
b) the emission is connected with current multipoles instead of mass multipoles.

The amplitude of the mode is small at the beginning, but it increases in a timescale
of about 100 s, until a non-linear regime is reached and when  saturation effects occur.
This saturation phase is the most likely to be detected and lasts about  10$^4$-10$^5$ s for
a crusted star and about 10$^6$-10$^7$ s for a fluid star. The expected strain amplitude 
in the saturation phase, calculated for a polytropic equation of state with n=1 and 
for a canonical neutron star of 1.4 M$_{\odot}$ is (Andersson \& Kokkotas 2000)
\begin{equation}
h(t) \approx 9.25\times 10^{-25}\alpha({{20 Mpc}\over{r}})
\end{equation}
where $\alpha$ is the dimensionless radial amplitude of the mode and r is the distance
to the source. In order to estimate the expected
S/N ratio the usual matched-filtering approach is adopted, although it must be
recognized that  such a tecnique is unlikely to be possible for this kind of signal. In
this case, the S/N ratio is given by
\begin{equation}
({{S}\over{N}})^2 = 4\int^{\infty}_0({{h_c}\over{h_n}})^2dln\nu
\end{equation}
where h$_n$ = $\lbrack \nu S_n(\nu) \rbrack ^{1/2}$, $S_n(\nu)$ is the noise 
spectral density and h$_c$  is the characteristic amplitude defined by
\begin{equation}
h_c = h(t)\lbrack \nu^2\mid {{dt}\over{d\nu}}\mid \rbrack^{1/2}
\end{equation}
This last relation is a consequence of the stationary phase approximation, meaning
that the detectability of a quasi-periodic signal is improved as the square root of the
number of cycles at a given frequency $\nu$. Using the planned VIRGO
sensitivity curve one obtains for a source at a distance of 20 Mpc,  S/N $\approx$ 2.2$\alpha$
if no crust is present and S/N $\approx$ 1.5$\alpha$ if a crust is already present in the star.
These S/N ratios are smaller by a factor 2-3 for LIGO and almost one order of magnitude
higher for the planned advanced LIGO. These estimates indicate that if the mode amplitude 
at saturation is nearly unity, then fast rotating newly born neutron stars  could be
good source candidates for the present generation of laser beam interferometers.

\subsection{Oscillating neutron stars}

Neutron stars have a large number of families of distinct pulsating modes (see
Kokkotas 1997 for a review). Nonradial oscillating modes as an emission
mechanism of GW were already discussed in the late sixties (Thorne \& Campolattaro
1967; Thorne 1969). These early calculations were concentrated mainly in the
so-called {\it f} (fundamental) mode, since this is the mode through which most of
the mechanical energy of the star is radiated away. All the other possible fluid modes
as {\it g} (gravity), {\it p} (pressure), {\it s} (shear), {\it t} (toroidal) and {\it i}
(interface) can
be calculated with sufficient accuracy using Newtonian dynamics, since they  don't
emit significant amounts of gravitational radiation.

The eigenfunctions of the f-modes have no nodes inside the star, reaching a maximum
at the surface of the star. Lindblom \& Detweiler (1983)  calculated quadrupole f-modes 
for a series of neutron stars characterized by different equations of state (see 
also Kokkotas \& Schutz 1992). Frequencies are in the range 1.5-4.0 kHz and the damping
timescales are of the order of  few tenths of second. In a first approximation, f-mode
frequencies and damping timescales can be estimated from the formulae
\begin{equation}
\nu_{kHz} = 46.13M^{0.255}R^{-1.32}
\end{equation}
and
\begin{equation}
\tau_{sec} = 5.164\times10^{-3}R^{1.60}M^{-0.813}
\end{equation}
where M and R are in solar units and km respectively.

As mentioned above, other oscillation modes are present in a real star. Pressure is the
restoring force for p-modes and frequencies associated to these modes depend on the
travel time for acoustic waves to cross the star. These frequencies are higher than 5 kHz
and the oscillations are damped in timescales longer than those of f-modes. The g-modes,
restored by gravity, depend critically on the internal composition and temperature 
profile, having  frequencies tipically of the order of few hundred Hz. The interplay of
all these modes in a neutron star is quite complex, since these objects have a solid 
crust and a central fluid interior. The modes f, p and g belong to the class of polar modes,
but if the shear modulus in the crust is non-zero, axial modes should exist as well as
families (i-modes) associated with the interface between distinct phases of the
matter inside neutron stars (McDermott et al. 1988). Besides these  ''Newtonian-modes'', 
there is a class of modes
uniquely associated with perturbations of the spacetime itself, the so-called
{\it w}-mode (Kojima 1988; Kokkotas \& Schultz 1992). These gravitational {\bf w}ave
modes arise because of the trapping of GW by the spacetime curvature generated by the
background density. The w-modes exist for both polar and axial perturbations since they
do not depend on perturbations of a fluid. These modes have high frequencies
(above 7 kHz) and are damped in timescales shorter than one millisecond. They are
probably the natural way to recover any initial deformation of the spacetime, as those
expected to occur during the gravitational collapse, leading to black hole formation.

One of the problems concerning the  gravitational wave emission from oscillating
neutron stars  is the absence of a convincing mechanism able to excite those modes.
Once the solid crust is formed, stresses will exist  by the presence of a strong
magnetic field or/and rotation. These stresses induce tectonic activity and
the stored elastic energy may be released as a consequence of quakes.
The elastic energy  is converted into shear waves that excite nonradial 
oscillation modes damped by GW. However, the maximum
elastic energy that can be channeled into oscillating modes is likely to be about 10$^{45}$ 
erg, restricting considerably the detection of gravitational wave sources excited by
tectonic activity.

The maximum distance that a given source can be detected by a gravitational wave
antenna can be estimated by the following considerations.
After filtering the signal, the expected S/N ratio is given by the equation
\begin{equation}
{(S/N)^2} = 4\int^{\infty}_0{{\mid \bar h(\nu)\mid ^2}\over{S_n(v)}}d\nu
\end{equation}
where $ \bar h(\nu)$ is the Fourier transform of the signal, here a sinusoidal damped
oscillation of amplitude h$_0$, and  $S_n(\nu)$ is the noise power spectrum of the
detector. Performing the required calculations, the S/N ratio can be written as
\begin{equation}
{(S/N)^2} = {4\over 5}{h_0^2}({{\tau_f}\over{S_n(\nu_0)}}){{Q^2}\over{1 + 4Q^2}}
\end{equation}
where $\tau_f$ is the damping time, $Q = \pi\nu_0\tau_f$ is the quality factor of the
oscillation and $\nu_0$ is the frequency of the considered mode. In the above equation, the
angle averaged beam factors of the detector were already included. The amplitude
of the signal is related with the total released energy E by the equation
\begin{equation}
{h_0} = {{2}\over{\pi\nu_0r}}\lbrack{{GE}\over{\tau_fc^3}}\rbrack ^{1/2}
\end{equation}
where r is again the distance to the source. From
these equations, once the energy and the S/N ratio are fixed, 
the maximum distance to a given source that a gravitational antenna can probe
can be estimated.
Using the planned sensitivities of VIRGO and LIGO as well as the oscillation
properties of neutron star models calculated by Lindblom \& Detweiler (1983), 
it is clear that sources
excited by starquakes cannot be detected beyond distances 
of about 2.5 kpc (de Freitas Pacheco 1998).

Another possible excitation mechanism of nonradial oscillations is the micro-collapse
suffered by a neutron star when a phase transition occurs in the core, caused
by a softening of the equation of state. There are
presently several arguments in favor of a stiff equation of state for the nuclear matter:

Firstly, high frequency  quasi-periodic oscillations (QPOs) have been found in the
X-ray emission originated from the accretion disk around neutron stars associated
to low-mass binaries (see van der Klis 2000, for a review). These oscillations are
observed in the range 300-1300 Hz, and are often splitted into pairs with a nearly
constant separation of 250-350 Hz. A possible interpretation of such a modulation
in the X-ray emission is the presence of instabilities (''hot spots'') near 
the  ''last stable'' orbit 
($r = {{6GM}\over{c^2}}$) and modulated by the local orbital frequency. In 
this case, the mass of the neutron star is directly related with the orbital frequency by
\begin{equation}
M = {1\over{(216)^{1/2}}}{{c^3}\over{G\Omega_{orb}}}\approx 2.2\nu_{kHz}^{-1} \,\  
  M_{\odot}
\end{equation}
If this interpretation is correct, neutron stars with masses up to 2.0 M$_{\odot}$ are
present in those low-mass X-ray binaries as a consequence of the accretion process. Moreover,
the analysis of the orbital motion of some massive X-ray binaries seems also to
suggest the presence of massive neutron stars (for instance, 1.8 M$_{\odot}$ for
4U1700-37 and 1.75 M$_{\odot}$ for Vela X-1). These
high mass values favor stiff equations of state (Akmal et al. 1998). Secondly, post-glitch 
recovery analyzes of isolated pulsars are not consistent with soft
equations of state (Link et al. 1992; Alpar et al. 1993)

Masses of neutron stars derived from NS-NS binaries cluster around 1.4 M$_{\odot}$. 
Neutron star models of 1.4 M$_{\odot}$ built with stiff and moderately stiff equations
of state suggest central densities of about (4-5)n$_0$ (de Freitas Pacheco et al. 1993;
Akmal et al. 1998), where n$_0$ = 0.16 fm$^{-3}$ is
the nuclear  saturation  density.  These densities are lower than the critical 
density for kaon condensation, if 
kaon-nucleon  and nucleon-nucleon correlations are taken into account
(Pandharipande et al. 1995). In this case, according to those calculations, the
kaon condensation should only occur for densities around (6-7)n$_0$. The appearance of
a kaon condensate will soften the equation of state reducing the maximum stable
mass, yielding the star unstable (however, there are no evidences in favor of 
the presence of black holes in those low-mass X-ray binary systems!).
It is worth mentioning that the
deconfinement of cold matter is expected to occur around (7-9)n$_0$ (de Freitas Pacheco
et al. 1993). If the neutron star accretes mass, the central density increases
and may eventually to overcome the critical density necessary to occur
a phase transition. Further increase of the neutron star mass leads the star into a metastable
situation, until the occurrence of a structural transition of the whole star into
a new configuration of  minimum energy. Consider, for instance, the
appearance of quark matter in the core. Since the energy density of the quark
matter is higher than that of the hadron matter, the star must to contract 
(in a dynamical timescale)
 in order to extract gravitational energy, which will provide the  ''latent'' heat
of the phase transition.  In the  ''metastable'' hadron state, a 1.81 M$_{\odot}$ 
neutron star has a central density n$_c$ = 7.6n$_0$ and a radius R = 9.9127 km,
while after the micro-collapse the radius is reduced
by {\bf 5.2 m} and a quark core matter of 0.44 km radius is formed. The event
 releases a total gravitational energy of about 4.7$\times$ 10$^{50}$ erg, from
which 85\%  go to the phase transition and,  about  7$\times$ 10$^{49}$ erg are
converted into heat and/or into mechanical energy (de Freitas
Pacheco 1999). In the ideal case, if the latter amount of energy is channeled into
the quadrupole f-mode (the coupling with rotation favors the excitation of
nonradial modes) , the emitted GW from the considered neutron star model
will have a frequency of 2.6 kHz and
the corresponding damping timescale will be $\tau_f$ = 0.124 s. Using
equations (18) and (19), the maximum distance that these signals can be
detected by the present generation of interferometers is about 85 kpc, which
includes all sources in the Milky Way and in the Magellanic Clouds.   

\section{Gravitational Waves from Black Holes}

Black holes are expected to exist in our universe with masses ranging from a few
solar masses up to 10$^{10}$ M$_{\odot}$. Stellar mass black holes may  be 
formed in the core collapse of massive stars, by accretion either of a massive  neutron 
star or a small hole, by merging of two  neutron stars at the end-point of their
inspiral in a binary system, whereas very massive holes, probably present in a
considerable number of galactic nuclei, can be formed by different routes. 

Recent 2D-hydrodynamic simulations of core collapse (Fryer 1999) for a large
range of progenitor masses, indicate that partial fallback of the envelope drives
the compact core to collapse into a black hole. This occurs for progenitors
with masses M$>$ 20 M$_{\odot}$ while progenitors more massive than 40
M$_{\odot}$ form black holes directly. In spite all the uncertainties present
in those simulations, they permit nevertheless an estimate of the mass 
distribution and of the formation rate of these objects.

Newly formed black holes are likely to be quite  ''deformed''. They need
to radiate away energy in order to settle down into a quiescent and axisymmetric
state characterized only by their mass M and angular momentum J (a Kerr black hole).
Detailed calculations suggest that black hole oscillations are easy to trigger and
that quadrupole modes dominate the emission  (Stark \& Piran 1985a, 1985b). The
waveform of the initial burst depends on the details of the collapse, but the
late-time behavior  (the ''ring-down'' phase) has a well established damped
oscillatory form, which is essentially a function of  M and J (see, for instance,
Echeverria 1989; Finn 1992). For the mode l = m = 2, using the results of Echeverria
(1989), the frequency and the damping timescale are respectively estimated from
\begin{equation}
{\nu_{gw}} \approx 12({{M_{\odot}}\over{M_{bh}}})F(a)  \,\  kHz
\end{equation}
and 
\begin{equation}
{\tau_{gw}} \approx 5.55\times 10^{-5}({{M_{bh}}\over{M_{\odot}}})K(a)  \,\  s
\end{equation} 
where a = ${{Jc}\over{GM^2}}$ and useful fitting formulae for the functions F(a),
K(a) are
\begin{equation}
F(a) \approx {{100}\over{37}}-{{63}\over{37}}(1-a)^{0.3} \,\ and \,\ K(a) \approx
(1-a)^{-0.45}
\end{equation}
The amplitude of the signal can be written as
\begin{equation}
{h_0} = {{2.3\times 10^{-21}}\over{D_{Mpc}\surd(F(a)Q)}}
({{\varepsilon}\over{10^{-4}}})^{1/2}({{M_{bh}}\over{M_{\odot}}})
\end{equation}
where Q is again the quality factor of the oscillation, D$_{Mpc}$ is the distance to
the source in megaparsec and the radiation efficiency was defined such as
E = $\varepsilon M_{bh}c^2$ be the total gravitational wave energy emitted by the
source. Equations (24) and (18) permit an estimate of the maximum
distance that a given event could be detected, once the S/N ratio, the radiation 
efficiency, the mass and the angular momentum of the hole are fixed. Here, 
estimates were performed for a typical black hole mass of 9.0 M$_{\odot}$,
a radiation efficiency $\varepsilon = 10^{-4}$ and
a S/N ratio equal to two. The results are given in table 1 for two values of the
angular momentum parameter {\bf a}. 

\begin{table*}
\caption[1]{Maximum Probed Distances (Mpc)}
\begin{flushleft}
\begin{tabular}{lcccccccccccc}
\noalign{\smallskip}
\hline
\noalign{\smallskip}
a&Q&$\nu_{gw}$(kHz)&VIRGO&LIGO&LIGO-ad \\
\noalign{\smallskip}
\hline
\noalign{\smallskip}
0.2&2.57&1.48&1.25&0.43&83 \\
0.9&10.9&2.46&0.44&0.16&9 \\
\noalign{\smallskip}
\hline
\end{tabular}
\end{flushleft}
\end{table*}

Inspection of table 1 shows that the detection of slow rotating holes 
are more favorable, in 
spite of fast rotating holes have a higher oscillation quality 
factor. The reason is that the 
fundamental quadrupole frequency increases for higher angular momentum values  and
the sensitivity of most interferometers decreases for $\nu >$ 1 kHz. A serious  handicap
to detect these events
is that the signal is completely damped out after only a few cycles and could easily  be
confounded with transient disturbances produced, for instance, by the suspension of
the mirrors.  Moreover, the expected maximum distances for the 
present planned sensitivity of
VIRGO and LIGO imply quite small detection rates and, only for the advanced LIGO
a high event rate should be expected, namely, about one event/month under the assumption
that all stars with M$\geq$ 40 M$_{\odot}$ produce black holes.

\section{Binary Systems}

The gravitational wave emission resulting from the merger of two compact 
stars (NS-NS, NS-BH, BH-BH)
is a very attractive possibility due to the huge energy power implied in
the process. This mechanism was already discussed in the late seventies
by Clark et al. (1979), who suggested that gravitational antennas could be
able to probe the universe up to distances of about 200 Mpc. Here, the
case of a binary system constituted by two neutron stars (NS-NS system)
will be discussed in more detail.   

In the late inspiral phase, which precedes the final coalescence, relativistic
effects or post-Newtonian (PN) corrections are necessary in order to describe
the wave form and the amplitude of the signal (Damour \& Deruelle 1981;
Blanchet \& Sathyaprakash 1994, 1995). The main features of the GW emission
during the inspiral motion are an increasing radiated power 
and a wave frequency equal to twice the orbital frequency. The
"last" stable orbit imposes a limit to the maximum wave frequency (see
eq. (20), with M now being the total mass of the system). For a binary system constituted by two "canonical" neutron
stars (1.4 M$_{\odot}$) such a limit frequency is about 1.6 kHz. As a
consequence, during the epoch when the GW emission is in the bandwidth where
the sensitivity of VIRGO or LIGO is the greatest, the motion is still well
described by the Newtonian picture, whereas PN corrections become important
at frequencies where the sensitivity of those detectors is already considerably reduced.

After averaging both polarization components with respect to the inclination
angle of the orbital plane, the Fourier transform of the signal is
\begin{equation}
{\mid \bar h(\nu)\mid^2} = <\mid \bar h_+(\nu)\mid^2 + \mid \bar h_X(\nu)\mid^2> =
{1\over{12}}{({{G^5M^2}\over{\pi^4c^9}})^{1/3}}{{\mu}\over{r^2\nu^{7/3}}}
\end{equation}
where M is the total mass of the system and $\mu$ is the reduced mass. 
Combining eqs.(17) and (25), the maximum distance probed by the
detector can be estimated by assuming M = 2.8 M$_{\odot}$ and S/N = 2.
For VIRGO, r$_{max}$ = 46 Mpc and for LIGO, r$_{max}$ = 36 Mpc, if the
planned sensitivity of these detectors is used.

The maximum distance permits to evaluate the expected event rate if
the coalescence rate is known. Regimbau (private communication) has
recently reviewed the formation rate of NS-NS systems.
She assumed an initial binary system constituted by progenitors with
masses greater than 9 M$_{\odot}$. The more massive evolves faster,
explodes and produces a neutron star. Then, as the less massive star evolves it loses
mass, affecting the rotation period of the first formed pulsar. The
evolving star becomes a "He-star" and explodes probably as a type Ib
supernovae. The system is supposed to remain bounded after both explosions.
If evolution of the newly formed pulsar is not
affected by external torques other than the canonical magnetic dipole,
then its rotation period will have a secular increase similar to that
observed for "isolated" pulsars. Table 2 summarizes the present census
of NS-NS systems. The first column identifies the pulsar and the others
give respectively the rotation period in ms, the logarithm of the period
derivative, the orbital period in days, the orbital eccentricity and the
total mass of the system. Notice that for B1820-11 only a lower limit
for the total mass is known since the rate of the relativistic advance 
of the longitude of the periastron is poorly determined. All 
the systems, excepting  PSR B1820-11, seem to
contain recycled pulsars, indicating that the first formed neutron star
is the one being observed. On the other hand, in the case of B1820-11, the
period derivative, the magnetic field and the indicative age favor
the interpretation of a young and non-recycled pulsar. However, the nature
of the companion is not yet well established. Lyne \& McKenna (1989)
suggested that the system is indeed constituted by two neutron stars, but
alternative possibilities like a main-sequence star (Phinney \& Verbunt
1991) or a white dwarf (Portegies Zwart \& Yungelson 1999) have also been
discussed in the literature. According to the simulations by Regimbau, in
order to observe one system like B1820-11, the present formation rate
R$_b(T)$ of NS-NS
binaries must be equal to 1.2$\times 10^{-5}$ yr$^{-1}$, consistent with
other recent estimates (Kalogera \& Lorimer 2000).

\begin{table*}
\caption[2]{NS-NS Binary Systems}
\begin{flushleft}
\begin{tabular}{lcccccccccccc}
\noalign{\smallskip}
\hline
\noalign{\smallskip}
PSR& P (ms)&log($\dot P$)&P$_{orb}$ (days)&e&M/M$_{\odot}$ \\
\noalign{\smallskip}
\hline
\noalign{\smallskip}
B1913+16&59.03&-17.10&0.323&0.617&2.83\\
J1518+49&40.94&-19.40&8.634&0.248&2.62\\
B1534+12&37.90&-17.60&0.420&0.274&2.68\\
B2127+11C&30.53&-17.30&0.335&0.681&2.71\\
J1811-17&104.18&-17.74&18.779&0.828&2.60\\
B1820-11&279.83&-14.86&357.76&0.795&$>$0.54\\
\noalign{\smallskip}
\hline
\end{tabular}
\end{flushleft}
\end{table*}

The coalescence rate can now be calculated following the procedure by
de Freitas Pacheco (1997). The first step is to compute the fraction $\gamma_b$
of formed stars giving rise to NS-NS systems. The formation rate of
NS-NS systems can be written as
\begin{equation}
R_b(t) = \gamma_bkM_g(t)\int_{m_1}^{m_2}\xi(m)dm
\end{equation}
where kM$_g(t)$ is the star formation rate (k in yr$^{-1}$ is the star
formation efficiency and M$_g(t)$ is the gas mass in solar masses in the Galaxy), 
$\xi(m)$ is the initial mass function and m$_1$ = 0.1M$_{\odot}$, 
m$_2$ = 80M$_{\odot}$ are respectively the minimum and the maximum stellar
masses. Adopting a galactic age T=15 Gyr and using the NS-NS formation
rate estimated above, from eq.(26) one obtains for the fraction of formed NS-NS binaries
$\gamma_b$ = 2.6$\times 10^{-6}$.
Then, assuming $\gamma_b$  constant in time, the second step consists  to
compute the coalescence rate $\varpi$ from the equation (de Freitas Pacheco 1997)
\begin{equation}
\varpi(t) = \int_{t_0}^t R_b(t-t')\phi(t')dt'
\end{equation}
where $\phi(\tau)={{B}\over{\tau}}$ is the probability for a NS-NS system to merge
in a timescale $\tau$ and t$_0$ is the minimum time required for a given pair
to coalesce. Solving the above integral, the resulting  coalescence
rate in the Galaxy at present is $\varpi(T) = 3.0\times 10^{-5} \,\ yr^{-1}$. This
procedure can now be applied to an elliptical galaxy, assuming the same
value of $\gamma_b$, but a higher efficiency for the star formation rate. Then,
a ''cosmic'' mean coalescence rate can be estimated by using a weighted
contribution of 20\% of ellipticals and 80\% of spiral galaxies. These
calculations give $\bar \varpi$ = 7.0$\times 10^{-7}$ Mpc$^{-3}$yr$^{-1}$, from
which an event rate of one merging every 3.5 yr is expected for VIRGO and one
every 7.3 yr for LIGO. Clearly, reasonable event rates will only be obtained with
the  ''enhanced'' versions of these detectors.
However, it is worth mentioning  that already a gain of a factor of two in the 
sensitivity implies an increase of one order of magnitude in the probed volume 
and the consequences of such an achievement may drastically change the 
situation. It seems to be rather well established now that at a distance of 
about 60 Mpc, in the direction of Centaurus, there is a huge concentration 
of galaxies dubbed the Great Attractor (Burstein 1990). The estimated mass of 
this supercluster is about 5$\times 10^{16}$  M$_{\odot}$ (but see Staveley-Smith 
et al. 2000), implying a coalescence rate of 11 events/yr from the
Great Attractor alone. However, we should have in mind the uncertainties
still present in all these estimates, once they are based on rather poor
statistics of NS-NS pairs.    

\section{The Stochastic and the Cosmological Background}

The stochastic background is the GW emission arising from a very large number of
unresolved and uncorrelated sources. A quantity which is often used 
to measure whether the collective effect of bursts generates a continuous 
background or not, is the so-called duty cycle defined by
\begin{equation}
D(z) = \int^z_0 \tau_*(1+z')dR_*(z')
\end{equation}
where $\tau_*$ is the average duration of a single burst at emission, R$_*(z)$ is
event rate at redshift z and the factor (1+z) takes into account the time dilation. If
D$<<$1, the background cannot be considered  as a continuous one, but should rather
be seen as a shot noise process. Supernovae (Blair et al.1997) and distorted stellar
black-holes (Ferrari et al. 1999a; de Ara\'ujo et al. 2000) are examples of sources
which may produce a shot noise like background. On the opposite situation, when
D$>>$1 a truly continuous background will be produced, as that due to
the r-mode emission from neutron stars (Owen et al. 1998; Ferrari et al. 1999b) or
from processes that took place very shortly after the big-bang.

In order to compare the importance of the GW background with other forms of energy
which dominate the expansion of the universe, it is usual to define an equivalent
energy density parameter $\Omega_{gw}$ associated to GW by the relation
\begin{equation}
\Omega_{gw}(\nu) =  {{1}\over{\rho_c}}({{d\rho_{gw}}\over{dln\nu}})
\end{equation}
where ${\rho_c}$ = ${{3c^3H_0^2}\over{8\pi G}}$ is the critical density and
$\rho_{gw}$ is the energy density under the form of gravitational waves.

Primordial nucleosynthesis and millisecond pulsars impose relevant constraints
on the closure density due to GW. The outcome from primordial nucleosynthesis
depends on the balance between the nuclear reaction rates and the expansion
of the universe. Thus, in order not to spoil the predict abundances of 
$^2$H, $^3$He, $^4$He and $^7$Li, which are in agreement with present data,
the energy density under the form of GW cannot exceed a certain limit. This
bound is usually established in terms of an effective number of neutrino
species N$_{\nu}$ and is given by (Kolb \& Turner 1990)
\begin{equation}
\int \Omega_{gw}(\nu)dln\nu \leq 1.3\times 10^{-5}(N_{\nu}-3)
\end{equation}
if the Hubble expansion parameter H$_0$ is taken to be equal to 65 km/s/Mpc. Primordial
nucleosynthesis constrains the number of massless neutrinos to be
N$_{\nu} \leq$ 3.2, restricting the background spectrum of GW over a wide
range of frequencies to have an equivalent amplitude $\Omega_{gw} < 10^{-6}$.
Notice that this is not a very restrictive constraint. 

Millisecond pulsars are very precise and stable clocks. The regularity of
the pulses can be described in terms of timing residuals, which represent
the errors in predicting the arrival time of these pulses. GW passing
between the pulsar and the observer cause a fluctuation in the arrival
time proportional to the amplitude of the perturbing waves. Eight years
of monitoring of PSR B1855+09 give an upper limit at $\nu \sim 10^{-8}$
Hz of $\Omega_{gw} < 10^{-8}$.

It is interesting to compare these upper bounds with the expected sensitivity
of laser beam interferometers. At lower frequencies, $\nu \sim$ 1 mHz,
LISA will be able to detect after one year integration, a GW background 
corresponding to a closure density $\Omega_{gw} \sim 10^{-12}$. The present
generation of ground based interferometers like VIRGO and LIGO, after one
year integration and using signal correlation techniques, may detect
a GW background equivalent to $\Omega_{gw} \sim 10^{-5}$, while the next
generation will be able to detect at frequencies in the range 0.1 - 1.0 kHz,
amplitudes several orders of magnitude smaller, namely, $\Omega_{gw} \sim
10^{-10}$ (Maggiore 2000).

The GW emission from core collapse (see section 2.2) is likely
to be of the order of 10$^{-8}$ Mc$^2$. Thus, the expected contribution
to the stochastic background will be about $\Omega_{gw} \leq 10^{-15}$, a
extremely small value in comparison with the expected amplitude from other 
sources. The GW stochastic
background due to black hole formation after the core collapse of a massive
star was estimated by Ferrari et al. (1999a) and de Ara\'ujo et al. (2000). 
Both studies conclude that the spectral energy distribution attains a maximum
around 1 kHz with an amplitude corresponding to a closure density of 
about $\Omega_{gw} \approx 10^{-10}$. This result indicates that such an emission
will not be detected by the present generation of laser beam interferometers but
it is within the capabilities of the advanced versions. The GW background
resulting from the r-mode instability in young and rapidly rotating neutron stars
may be a more promising possibility. According to the computations by
Ferrari et al. (1999b), the closure density has a maximum amplitude plateau
of $\Omega_{gw} \sim 7\times 10^{-8}$ in the frequency range 0.5 - 1.7 Hz, which
is about two orders of magnitude above the expected detection threshold for
the advanced interferometers.
 
Different physical processes which could have taken place in the early universe
and which produce a continuous GW background, have been discussed in the literature.
The first mechanism consists on GW generated by quantum perturbations in a 
inflationary scenario. As the universe cools, it pass through a phase dominated
by the vacuum energy, where the scale factor increases exponentially, followed
by a rapid transition to a radiation dominated phase. GW quanta are produced
by quantum fluctuations during such a transition and the resulting closure 
density is (Allen 1997)
\begin{equation}
{\Omega_{gw}} = {{16\hbar G^2}\over{9c^7}}{\rho_v}
\end{equation}
where $\rho_v$ is the energy density of the vacuum in the inflationary phase. The
spectrum is flat at frequencies seen by LISA or ground based interferometers 
like VIRGO and LIGO, and the expected amplitude is of the order of
$\Omega_{gw} \sim 10^{-13}$. This value is well below the expected detection 
limit of LISA and ground based advanced interferometers. Another scenario
was developed by Vilenkin (1985) (see also Vilenkin \& Shelard 1994), who
considered the GW background produced by a network of cosmic strings
formed during a symmetry breaking phase transition. This scenario offers a
situation more favorable for a future detection. The spectrum is almost
flat in the frequency range 10$^{-9}$-10$^9$ Hz with an amplitude
$\Omega_{gw} \sim 10^{-8}$. The existence of this possible background component 
can be confirmed or not by a continuous monitoring of
the timing residuals of millisecond pulsars, by future space observations as
well as by the next generation of laser interferometers. More recent
studies in this domain (Damour \& Vilenkin 2000) seem to indicate that the stochastic
background produced by strings is non-Gaussian, including occasional bursts
emanating from cusps, that stands above the mean amplitude. According to
those authors, the GW bursts might be accompanied by gamma-ray bursts, what
could be a decisive signature of such a mechanism.

A more fascinating possibility is related with theories which describe our
world as a (mem)brane embedded in a higher dimensional space, having the
purpose of solving the huge energy gap that separates the electroweak scale
from the Planck scale, the so-called hierarchy problem (Randall \& Sundrum 1999).
In this scenario, GW are produced by coherently excited radions (geometrical
degree of freedom controlling the size or  curvature of the extra dimensions) and
Nambu-Goldstone modes (Hogan 2000). According to estimates by Hogan (2000),
the typical GW frequency is $\nu \sim 10^{-4}$ Hz and the amplitude corresponds
to a closure density of about $\Omega_{gw} \sim 8\times 10^{-5}$. Notice that
this amplitude is higher than the bound imposed by primordial nucleosynthesis.

\section{Concluding Remarks}

The direct detection of GW will constitute an extraordinary scientific
accomplishment, giving further support to the General Relativity Theory, opening
a new window to explore the Universe. 

GW from oscillating neutron stars excited by core phase transitions, 
from unstable newly born proto-neutron stars (r-modes) or from the 
last phases of the coalescence of a NS-NS binary, will give an important 
insight on the properties of the nuclear matter in these objects. Moreover, 
the eventual detection of GW from the ''ringdown'' phase of a newly formed 
black hole will permit to probe directly their properties like mass and angular
momentum. The detection rate of these events, if estimated by the future advanced 
interferometers, will impose severe constraints on the last evolutionary
phases of massive stars.

The Universe is highly transparent to GW and the detection of a background will
probe the very early phases of the big-bang, when radiation and matter are still
strongly coupled. Modern unified theories can also be tested as well as the
spacetime structure of our world.

{}

\end{document}